\documentclass[]{article}

\usepackage{times}
\usepackage[T1]{fontenc}

\usepackage{graphics,graphicx}
\usepackage{esvect}
\usepackage{amsmath,amsthm,amssymb}
\usepackage{url}

\usepackage{thmtools}
\usepackage{hyperref}

\declaretheorem[style=definition]{example}
\declaretheorem[style=definition]{definition}
\declaretheorem[style=definition]{proposition}
\declaretheorem[style=definition]{theorem}
\renewcommand\thmcontinues[1]{Continued}

\newcommand{\eg}{e.g.}
\newcommand{\ie}{i.e.}
\newcommand{\viz}{viz.}
\newcommand{\cf}{cf.}
\newcommand{\st}{\ensuremath{\mbox{s.t.}}}

\newcommand{\T}{\ensuremath{T}}
\newcommand{\dT}{\ensuremath{\mathit{DT}}}
\newcommand{\C}{\ensuremath{C}}
\newcommand{\dC}{\ensuremath{\mathit{DC}}}
\newcommand{\W}{\ensuremath{W}}
\newcommand{\m}{\ensuremath{m}}

\newcommand{\realset}{\ensuremath{\mathbb{R}}}
\newcommand{\naturalset}{\ensuremath{\mathbb{N}}}

\newcommand{\tw}[2]{\ensuremath{{}_{#1}TW_{#2}}}
\newcommand{\conf}[2]{\ensuremath{{}_{#1}C_{#2}}}

\newcommand{\tuple}[1]{\ensuremath{\langle #1 \rangle}}
\newcommand{\belief}[1]{\ensuremath{b_{#1}}}
\newcommand{\disbelief}[1]{\ensuremath{d_{#1}}}
\newcommand{\uncertainty}[1]{\ensuremath{u_{#1}}}
\newcommand{\opinion}[1]{\ensuremath{\tuple{\belief{#1}, \disbelief{#1}, \uncertainty{#1}}}}

\newcommand{\optrustconf}[2]{\ensuremath{#1 \circ #2}}

\newcommand{\tri}[1]{\ensuremath{\overset{\triangle}{#1}}}
\newcommand{\ang}[1]{\ensuremath{\angle_{#1}}}

\newcommand{\versor}[1]{\ensuremath{\vv{e_{#1}}}}

\title{Context-dependent Trust Decisions with Subjective Logic}

\author{Federico Cerutti~~  Alice Toniolo~~ Nir Oren~~ Timothy J. Norman\\University of Aberdeen}

\date{May 20, 2013}

\begin{document} 

 
\maketitle 

\begin{abstract} \mbox{}
A decision procedure implemented over a computational trust mechanism aims to allow  for decisions to be made regarding whether some entity or information should be trusted. As recognised in the literature, trust is contextual, and we describe how such a context often translates into a confidence level which should be used to modify an underlying trust value. J{\o}sang's Subjective Logic has long been used in the trust domain, and we show that its operators are insufficient to address this problem. We therefore provide a decision-making approach about trust which also considers the notion of confidence (based on context) through the introduction of a new operator. In particular, we introduce general requirements that must be respected when combining  trustworthiness and  confidence degree, and demonstrate the soundness of our new operator with respect to these properties.
%
\end{abstract}

\vspace{1em}
\vspace{1em}

\section{Introduction} 

Trust forms the backbone of human societies, improving system robustness by restricting the actions of untrusted entities and the use of untrusted information. Trust has also been studied within multi-agent systems \cite{Sabater2005}, where a trust value is associated with different agents. This level of trust is utilised by others when selecting partners for interactions; distrusted agents will rarely be interacted with, reducing their influence over the system.

Trust mechanisms aim to compute a level of trust based on direct and second-hand interactions between agents. The latter, commonly referred to as \emph{reputation} information, is obtained from other agents which have interacted with the agent whose trust is being computed. Aspects of such systems that have been examined include how to minimise the damage caused by collusion between agents \cite{Haghpanah12prep}, the nature of reputation information \cite{josang12trust}, and examining trust in specific contexts and agent interaction configurations \cite{burnett12subdelegation}.

It has also been recognised that trust is often context dependent. For example, while you may trust a mechanic (in the context of fixing your car), you would not trust him to perform heart surgery on you (in the context of a medical problem). One common approach to dealing with context involves holding a separate trust rating for each possible context. However, computing trust in the presence of a new context then becomes difficult, if not impossible \cite{Delmotte1996,Rousseau1998,Sabater2005}. 
In this paper, we assume that some sort of metric can be used to determine a similarity measure between contexts, leading to some \emph{confidence} measure. This confidence measures how much weight should be placed in a trust rating in the original context. We express both confidence and trust using a Subjective Logic opinions \cite{Josang2001}, and then seek to compute a final trust value based on both the original trust value and the degree of confidence.

In the next section we provide a running example which further motivates our work. We also provide a brief overview of Subjective Logic (hereafter evaluated SL). Section \ref{sec:core-properties} describes several core properties that any combination of trust and confidence must comply with, thus deriving some requirements that have to be satisfied when combining a SL opinion with another one representing the confidence on the former. Section \ref{sec:comb-trustw-conf} shows that existing subject logic operators do not meet these properties and then derives a new operator which we prove to be compliant with the properties. Then we discuss related and future work in Section \ref{sec:disc-future-works}, and draw conclusions in Section \ref{sec:conclusions}.

\section{Background and Motivations} 
\label{sec:backgr-motiv}

In this work we  focus on trust relations where the truster ``depends'' (using the same terminology of \cite{Castelfranchi2010}) on a trustee. As a concrete example, we examine the case where a trustee is responsible  for providing some information to a truster, as exemplified by the following scenario.

\begin{example}[label=exa:scenario]
  \label{ex:scenario}
  Let $X$ be a military analyst who is collecting evidences in order to decide whether or not a specific area contains a certain type of weapon. In particular, he needs a datum \m{} from sensor $Y$, which has a history of failures and this affects $X$'s perceived trustworthiness on $Y$. Here, $X$ is the truster, and $Y$ the trustee.
\end{example}

In such a scenario, the degree of trustworthiness of $Y$ is normally computed from historical data (\cf{} \cite{Josang2004}). We  consider four extensions of the basic scenario, wherein additional contextual information provides different  degrees of confidence that $X$ should trust $Y$'s information. More precisely, $X$ can have high, low, uncertain or intermediate confidence in $Y$'s information.

We begin with a  scenario where $X$ has high confidence in their trust computation, as  they have  no reason to doubt that their original  (perceived) trustworthiness degree should not be the case.

\begin{example}[continues=exa:scenario]
  \textbf{High confidence}. Suppose that $X$ knows that  $Y$ was recently maintained, and  therefore has no reason to believe that $Y$ should not be trustworthy. At the same time, however, $Y$ has had a history of (random) failures, meaning that $X$ does not completely trust $Y$'s reports. In such a scenario, it seems reasonable that $X$ should compute a degree of trustworthiness directly from historical interactions, as they have no reason for believing that this computed degree of trust does not hold in such a situation.
\end{example}

This scenario can be contrasted with a low confidence scenario, where additional knowledge allows $X$ to identify a different context, affecting the perceived degree of trustworthiness of reports from $Y$. This is captured by the following scenario.

\begin{example}[continues=exa:scenario]
  \textbf{Low confidence}. Suppose that $X$ knows that $Y$'s failures usually occur after a sandstorm, and that the data of interest was obtained following such an event. This is a strong reason for believing that the data collected are incorrect, despite the degree of trustworthiness which has been derived from the past interactions with $Y$. In such a situation, confidence in the degree of trustworthiness of $Y$ will be very low.
\end{example}

A third limiting case occurs when there are reasons to believe that the current context does affect the trust computation, but no information is available regarding \emph{how} it is affected. This is exemplified by the following situation.

\begin{example}[continues=exa:scenario]
  \textbf{Uncertainty}. Consider the case where $Y$ is in an area under the control of the enemy. Since the enemy can act to deceive $Y$, $X$ might adopt a prudent approach and decide not to consider the information provided by $Y$. This is not because $X$ does not believe it, but rather because $Y$ might (but is not necessarily) being deceived by the enemy. Here, there is  complete uncertainty with regards to the degree of  trustworthiness associated with $Y$'s reports.
\end{example}

Finally, we can also identify an  intermediate case, where it is known that the current situation negatively affects the degree of trustworthiness, and where an estimate of this effect can be determined. This is illustrated in the following scenario.

\begin{example}[continues=exa:scenario]
\textbf{Intermediate confidence}. Suppose that $Y$ is known to operate well under certain environmental conditions, and that $X$ must decide how to act in a situation that violates these parameters\footnote{A similar example dealing with GPS data is described by \cite{Bisdikian2012}.}. In this case $X$ knows that the data they receive is somewhat accurate (given knowledge of $Y$'s behaviour), and can therefore have some confidence in their level of trust of $Y$.
\end{example}

In the above scenarios, we utilised two terms with a clear intuitive meaning, namely \emph{trustworthiness} and \emph{confidence}. These have well-established definitions in literature, which usually refer to the notion of \emph{trust}.

\begin{definition}[From \cite{Castelfranchi2010}]
  \emph{Trust} is a relation between:
  \begin{itemize}
  \item an agent $X$ (truster) which is a cognitive agent;
  \item an addressee $Y$ (trustee) which is an agent in the broader sense of this term;
  \item a casual process (act/performance) and its results, \viz{} an act $\alpha$ of $Y$ possibly producing an outcome $p$ desirable because it includes (or corresponds to) a goal of $X$;
  \item a goal of $X$;
  \item a context $C$ or situation or environment where $X$ takes into account $Y$ and/or where $Y$ is supposed to act.
  \end{itemize}

Below, using the notation introduced in \cite{Castelfranchi2010}, we abbreviate the trust relation as $\mathit{TRUST}(X, Y, C, \tau, g_X)$, meaning that  $X$ trusts (in the information provided by) $Y$ in the context $C$ for performing action $\alpha$ (executing task $\tau$) and realising the result $p$ that includes or corresponds to her goal $g_X$.
\end{definition}

One of the factors that determines the existence and the degree of this trust relation is the level or \emph{degree of trustworthiness} in $Y$ that $X$ is judged to have.

\begin{definition}[Adapted from \cite{Castelfranchi2010}]
  Given an agent $X$ (truster), an addressee $Y$ (trustee), a task $\tau$, a goal $g_X$ and a context $C$, the \emph{trustworthiness} of $Y$ is a property of $Y$ in relation to a potential partner $X$, concerning a task $\tau$ for reaching a goal $g_X$ in a context $C$: in symbols $\tw{X}{Y}$. If $\tw{X}{Y} = \dT$, $\dT \in \realset^n$ for some $n \in \naturalset$, $n > 0$, then $\dT$ is the \emph{degree of trustworthiness} of $Y$ in relation to a potential partner $X$.
\end{definition}

  As noted in \cite{Castelfranchi2010}, $\tw{X}{Y}$ is one of the bases of the $\mathit{TRUST}(X, Y, C, \tau, g_X)$ relation, but the latter cannot be reduced to the former. In \cite{Castelfranchi2010} several ingredients of trust are considered, but in this paper we concentrate on only two of the components that contribute to the relation, namely trustworthiness and confidence. While the other elements must also be considered, we believe that addressing these two components goes a long way in the computation of trust. Therefore, in this paper we aim only to determine a degree of trust given some perceived degree of trustworthiness in an information source together with a \emph{degree of confidence} in the context under which the information is provided.
  
\begin{definition}[Adapted from \cite{Urbano2013}]
  Given an agent $X$ (truster), an addressee $Y$ (trustee), a task $\tau$, a goal $g_X$, a context $C$, and the trustworthiness of $Y$ $\tw{X}{Y}$, the \emph{confidence} of $X$ in $\tw{X}{Y}$ ($\conf{X}{\tw{X}{Y}}$) represents the reliance of the trustworthiness in the context $C$. If $\conf{X}{\tw{X}{Y}} = \dC$, $\dC  \in \realset^n$ for some $n \in \naturalset$, $n > 0$ is the \emph{degree of confidence} of $X$ in $\tw{X}{Y}$.
\end{definition}



Following \cite{Josang2007}, we express both the degree of trustworthiness and the degree of confidence using Subjective Logic (SL). This formalism extends probability theory expressing uncertainty about the probability values themselves, which makes it useful for representing trust degrees. We now proceed to provide a brief overview of SL mainly based on \cite{Josang2001}.

Like Dempster-Shafer evidence theory \cite{Dempster1968,Shafer1976}, SL  operates on a
\emph{frame of discernment}, denoted by $\Theta$. A frame of discernment
contains the set of possible system states, only one of which represents the
actual system state. These are referred to as atomic, or primitive, system states.
The powerset of
$\Theta$, denoted by $2^\Theta$, consists of all possible unions of primitive
states. A non-primitive state may contain other states within it. These are
referred to as substates of the state.

\begin{definition} \label{beliefMassAssignment}
Given a frame of discernment $\Theta$, we can associate a belief mass
assignment $m_\Theta(x)$ with each substate $x \in 2^\Theta$ such that
\begin{enumerate}
  \item $m_\Theta(x) \geq 0$;
  \item $m_\Theta(\emptyset) = 0$;
  \item $\displaystyle\sum_{x \in 2^\Theta} m_\Theta(x)=1$.
\end{enumerate}
\end{definition}

For a substate $x$, $m_\Theta(x)$ is its \emph{belief mass}.

Belief mass is an unwieldy concept to work with. When we speak of belief in
a certain state, we refer not only to the belief mass in the state, but also
to the belief masses of the state's substates. Similarly, when we speak
about disbelief, that is, the total belief that a state is not true, we need
to take substates into account. Finally, SL also introduces the
concept of uncertainty, that is, the amount of belief that might be in a
superstate or a partially overlapping state. these concepts can be 
formalised as follows.

\begin{definition} Given a frame of
discernment $\Theta$ and a belief mass assignment $m_\Theta$ on $\Theta$, we
 define the belief function for a state $x$ as
$$b(x)=\sum_{y \subseteq x} m_\Theta(y) \textrm{ where } x,y \in 2^\Theta$$
The disbelief function as
$$d(x) = \sum_{y \cap x=\emptyset} m_\Theta(y) \textrm{ where } x,y \in
2^\Theta$$
And the uncertainty function as
$$u(x) = \sum_{\tiny{\begin{array}{l}y \cap x \neq \emptyset \\ y \not \subseteq x
\end{array}}} m_\Theta(y) \textrm{ where } x,y \in 2^\Theta$$
\end{definition}

These functions have two important properties. First, they
all range between zero and one. Second, they always sum to one, meaning that
it is possible to deduce the value of one function given the other two.




Boolean logic operators have SL equivalents. It makes sense to
use these equivalent operators in frames of discernment containing a state and
(some form of) the state's negation. A \emph{focused frame of discernment} is
a binary frame of discernment containing a state and its complement.

\begin{definition} Given $x \in
2^\Theta$, the frame of discernment denoted by $\tilde{\Theta}^x$, which contains
two atomic states, $x$ and $\neg x$, where  $\neg x$ is the complement of $x$
in $\Theta$, is the focused frame of discernment with focus on $x$. 

Let $\tilde{\Theta}^x$ be the focused frame of discernment with
focus on x of $\Theta$. Given a belief mass assignment $m_\Theta$ and the
belief, disbelief and uncertainty functions for $x$ ($b(x)$, $d(x)$ and $u(x)$
respectively), the focused belief mass assignment, $m_{\tilde{\Theta}^x}$ on
$\tilde{\Theta}^x$ is defined as
\begin{eqnarray*}
&& m_{\tilde{\Theta}^x}(x) = b(x) \\
&& m_{\tilde{\Theta}^x}(\neg x) = d(x) \\
&& m_{\tilde{\Theta}^x}(\tilde{\Theta}^x) = u(x) 
\end{eqnarray*}
The focused relative atomicity of $x$ (which approximates the role of a prior probability distribution within probability theory, weighting the likelihood of some outcomes over others) is defined as
$$a_{\tilde{\Theta}^x}(x/\Theta)=[E(x)-b(x)]/u(x)$$
For convenience, the focused relative atomicity of $x$ is often abbreviated $A_{\tilde{\Theta}^x}(x)$.
\end{definition}

An opinion consists of the belief, disbelief, uncertainty and relative
atomicity as computed over a focused frame of discernment.

\begin{definition}
Given a focused frame of discernment $\Theta$ containing $x$ and its complement
$\neg x$, and assuming a belief mass assignment $m_\Theta$ with belief,
disbelief, uncertainty and relative atomicity functions on $x$ in $\Theta$ of
b(x),d(x),u(x) and a(x), we define an \emph{opinion} over x, written $\omega_x$ as
$$\omega_x \equiv \langle b(x),d(x),u(x),a(x) \rangle$$
\end{definition}

For compactness, J{\o}sang also denotes the various functions as
$b_x$,$d_x$,$u_x$ and $a_x$ in place, and we will follow his notation. Furthermore, given a fixed $a_x$, an opinion $\omega$ can be denoted as a $\langle b_{x},d_{x},u_{x} \rangle$ triple.

Given opinions about two propositions from different frames of discernment, it
is possible to combine them in various ways using operators introduced, above all, in \cite{Josang2001,Josang2004,Josang2005,Josang2006,McAnally2004}. Reviewing these operators is beyond the scope of this paper, which instead is aimed at providing an approach for deciding whether or not to trust a source of information given a degree of trustworthiness and a degree of confidence.

\section{Core Properties and Requirements}
\label{sec:core-properties}

In the scenario described in Example \ref{ex:scenario}, an agent $X$ has to determine whether or not to trust a message \m{} received from $Y$. $X$ will consider three elements for reaching a decision:

\begin{enumerate}
\item \textbf{trustworthiness:} $X$ has an opinion \T{} concerning the degree of trustworthiness to assign to $Y$ considering the message \m; 

\item \textbf{confidence:} $X$ has an opinion \C{} about his own confidence regarding her opinion \T{} in this specific case;

\item \textbf{combination:} $X$ has to combine \T{} with \C{} in order to achieve an ultimate opinion \W{} on \m. This combination has to fulfil the following \emph{combination requirements}:
  \begin{enumerate}
  \item if \C{} is pure belief, then $\W = \T$;
  \item if \C{} is pure disbelief, then $\W = \C$ (i.e.~$\langle 0,1,0\rangle$);
  \item if \C{} is completely uncertain, then $\W = \C$ (i.e.~$\langle 0,0,1\rangle$);
  \item the degree of belief of \W{} is always less than or equal to the degree of belief of \T.
  \end{enumerate}

Note that the combination requirements describe the same ``prudent'' behaviour which has been presented in Example \ref{ex:scenario}, in particular in the ``high confidence'' scenario. Indeed even if $X$ is highly confident in a specific context, this confidence cannot increment the trust degree over the base trustworthiness degree.

\end{enumerate}

Following \cite{McAnally2004,Josang2006}, we utilise SL to instantiate trustworthiness and confidence, and seek to compute their combination through SL operations\footnote{Hereafter each opinion will have a fixed relative atomicity of $\frac{1}{2}$.}. In doing so, we must therefore consider the following inputs and requirements:

\begin{enumerate}
\item $\T = \opinion{\T}$ derived by statistical observations (\eg{} \cite{McAnally2004});
\item $\C = \opinion{\C}$ considering the specific context at hand;
\item $\W = \left\{ 
    \begin{array}{l l}
      \T & \mbox{if}~\C = \tuple{1, 0, 0}\\
      \C & \mbox{if}~\C = \tuple{0, 1, 0}\\
      \C & \mbox{if}~\C = \tuple{0, 0, 1}\\
    \end{array}
  \right.$

  further requiring that  $\belief{\W} \leq \belief{\T}$.
\end{enumerate}

Since 1.~and 2.~above are inputs, we concentrate on the constraints expressed by 3., which require us to consider the problem of how to combine the degree of trustworthiness with the degree of confidence. Existing work, such as \cite{McAnally2004,Castelfranchi2010,Urbano2013}, concentrate on computing $\T$. The problem of deriving the degree of confidence from a specific context (input 2.) is beyond the scope of this work but will form part of our future work.

\section{Combining Trustworthiness and Confidence}
\label{sec:comb-trustw-conf}

We begin by noting --- as illustrated in Table \ref{tab:comparison-josang-operators} --- that none of the operators provided by SL  \cite{Josang2001,Josang2004,Josang2005,Josang2006,McAnally2004,Josang2008} satisfy the combination requirements described previously.

\begin{table}[h]
  \centering
  \begin{tabular}{l | c | c | c | c }
                   & \multicolumn{4}{c}{Requirement}\\
    Operator       & (a) & (b) & (c) & (d)\\
    \hline
    Addition ($+$) & No & No & No & No \\
    Subtraction ($-$) & No & No & No & Yes \\
    Multiplication ($\cdot$) & No & Yes & No & No \\
    Division ($/$) & No & No & No & No \\
    Comultiplication ($\sqcup$) & No & No & No & No \\
    Codivision ($\sqcap$) & No & No & No & No \\
    Discounting ($\otimes$) & No & No & Yes & Yes \\
    Cumulative fusion ($\oplus$) & No & Yes & No & No \\
    Averaging fusion ($\underline{\oplus}$) & No & No & No & No \\
    Cumulative unfusion ($\ominus$) & No & Yes & No & No \\
    Averaging unfusion ($\underline{\ominus}$) & No & Yes & No & No \\
  \end{tabular}
  \caption{J{\o}sang operators and the satisfaction of the four combination requirements}
  \label{tab:comparison-josang-operators}
\end{table}

Therefore, it is necessary to define a new Subjective Logic operator which complies with our  combination requirements. To derive such an operator, we proceed in three steps.
\begin{enumerate}
\item We identify a legal set of opinions that satisfies combination requirement (d). This legal set has specific geometric properties, allowing us to project an opinion onto a bounded plane.
\item We must identify a specific projection method.
\item Finally the method chosen in step 2 is used to project the confidence opinion to  the set of values determined at step 1.
\end{enumerate}

In order to prove the soundness of our approach, we first need to discuss the geometry of Subjective Logic.

\subsection{The Geometry of Subjective Logic}
\label{sec:geom-subj-logic}

\begin{figure}[h]
  \centering
  \includegraphics[scale=0.7]{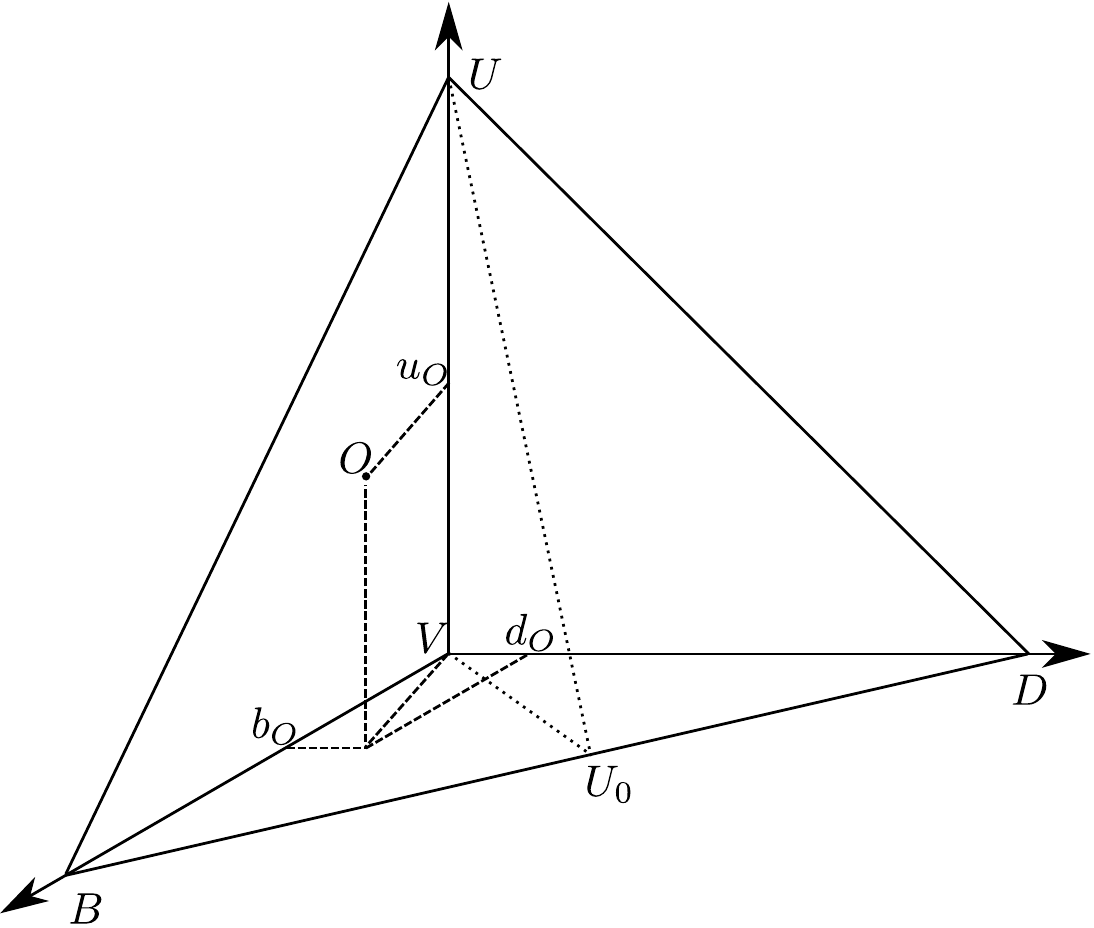}
  \caption{The Subjective Logic plane region}
  \label{fig:3d-sl}
\end{figure}

A SL opinion $O \triangleq \opinion{O}$ is a point in the $\realset^3$ space, identified by the coordinate \belief{O} for the first axis, \disbelief{O} for the second axis, and \uncertainty{O} for the third axis. However, due to the requirement that $\belief{O} + \disbelief{O} + \uncertainty{O} = 1$, an opinion is a point inside (or at least on the edges of) the triangle $\tri{BDU}$ shown in Figure \ref{fig:3d-sl}, where $B = \tuple{1, 0, 0}, D=\tuple{0,1,0}, U=\tuple{0,0,1}$.

\begin{definition}
  The \emph{Subjective Logic plane region \tri{BDU}} is the triangle whose vertices are the points $B \triangleq \tuple{1,0,0}$, $D\triangleq\tuple{0,1,0}$, and $U\triangleq\tuple{0,0,1}$ on a $\realset^3$ space where the axes are respectively the one of belief, disbelief, and uncertainty predicted by SL.
\end{definition}

Now $|\vv{VB}| = |\vv{VD}| = |\vv{VU}| = 1$ and therefore, since  angle $\ang{BVD} = \frac{\pi}{2} = \ang{VU_0B}$, $|\vv{BD}| = \sqrt{2}$, $|\vv{U_0B}| = \frac{\sqrt{2}}{2} = |\vv{VU_0}|$, it is the case that  $\vv{U_0U}=\frac{\sqrt{3}}{\sqrt{2}}$. Given that, the angle $\ang{UBD} = \arctan(\frac{|\vv{U_0U}|}{|\vv{U_OB}|}) = \frac{\pi}{3}$,  triangle $\tri{BDU}$ is equilateral where each slide is $\sqrt{2}$ and thus each altitude is $\frac{\sqrt{3}}{\sqrt{2}}$.

\begin{figure}[h]
  \centering
  \includegraphics[scale=0.7]{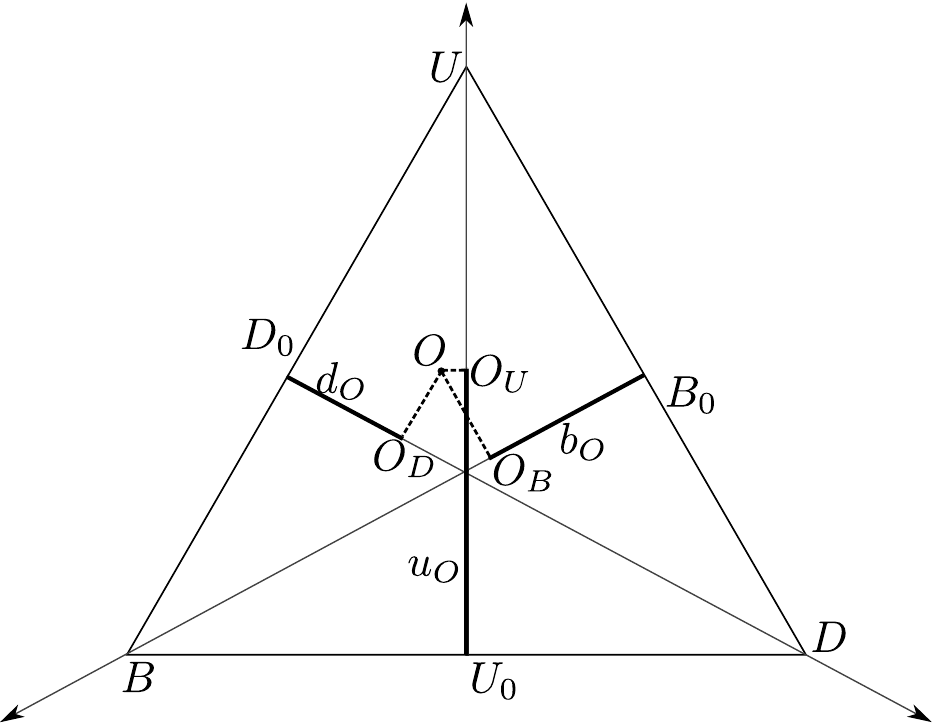}
  \caption{An opinion $O \triangleq \opinion{O}$ in SL after the $1:\frac{\sqrt{3}}{\sqrt{2}}$ scale. The belief axis is the line from $B_0$ (its origin) toward the $B$ vertex, the disbelief axis is the line from $D_0$ toward the $D$ vertex, and the uncertainty axis is the line from $U_0$ toward the $U$ vertex}
  \label{fig:basic-sl}
\end{figure}

Since each opinion is a point inside  triangle $\tri{BDU}$, it can be mapped to a point in Figure \ref{fig:basic-sl}. This representation is similar to the one used in \cite{Josang2001} for representing opinions in SL, but here the belief and disbelief axes are swapped for reasons that will become clear below.

In order to keep the discussion consistent with J{\o}sang's work \cite{Josang2001}, in what follows we will scale triangle \tri{BDU} by a factor $1:\frac{\sqrt{3}}{\sqrt{2}}$ thus obtaining that $|\vv{B_0 B}| = |\vv{D_0 D}| = |\vv{U_0 U}| = 1$. For convenience, we write the following.

\begin{enumerate}
\item $\displaystyle{\frac{1}{|\vv{B_0 B}|} \vv{B_0 B}} = \versor{\belief{}}$ as the unit vector of the axis of belief;
\item $\displaystyle{\frac{1}{|\vv{D_0 D}|} \vv{D_0 D}} = \versor{\disbelief{}}$ as the unit vector of the axis of disbelief;
\item $\displaystyle{\frac{1}{|\vv{U_0 U}|} \vv{U_0 U}} = \versor{\uncertainty{}}$ as the unit vector of the axis of uncertainty.
\end{enumerate}

More precisely however, $\displaystyle{\frac{1}{|\vv{U_0 U}|} \vv{U_0 U}}$ is the unit vector of the projection of the unit vector of the axis of uncertainty on a plane which intersects the plane determined by the Cartesian product of the unit vector of the axis of belief and the unit vector of the axis of disbelief; the points $\tuple{1, 0, 0}$ and $\tuple{0, 1, 0}$ are in that intersection, and the angle formed by the two planes is $\arctan(\frac{|\vv{VU}|}{|\vv{VU_0}|}) = \arctan(\sqrt{2}) = \varphi$ (according to an observer on $V$). Therefore, a point $\displaystyle{\frac{p}{|\vv{V U}|} \vv{V U}}$ (a point on the axis of uncertainty distant $p$ from $V$) can be projected into the triangle $\tri{BDU}$ using a line parallel to the plane determined by the Cartesian product of the unit vector of the axis of belief and the unit vector of the axis of disbelief, and the result of the projection will be the point $\displaystyle{\frac{\bar{p}}{|\vv{U_0 U}|} \vv{U_0 U}}$ where $\bar{p} = \frac{p}{\sin(\varphi)} = p~ \left(\frac{\tan(\varphi)}{\sqrt{1 + \tan^2(\varphi)}}\right)^{-1} = p~\left(\frac{\sqrt{2}}{\sqrt{3}}\right)^{-1} = p~\frac{\sqrt{3}}{\sqrt{2}}$. The same line of reasoning line can be applied to the remaining two axes.
Given the fact that the projection introduces a simple scale factor, we ignore it below in order to both simplify our discussion and maintain consistency with existing work.

With the above in hand,  given an opinion $O$, depicted in Fig. \ref{fig:basic-sl}, we can determine $\belief{O}$ as the distance ($|\vv{B_0O_B}|$) from the origin of the axis of beliefs ($B_0$) and the projection of $O$ on the same axis ($O_B$), where the projection is orthogonal ($\ang{OO_BB_0} = \frac{\pi}{2}$). The projection of $O$ on the axes of disbelief and of uncertainty will determine respectively $\disbelief{O}$ and $\uncertainty{O}$.

These geometric relations lie at the heart  of the Cartesian transformation operator which is the subject of the next subsection.

\subsection{The Cartesian Representation of Opinions}

As shown in \ref{sec:geom-subj-logic}, an opinion in SL can be represented as a point in a planar figure (Fig. \ref{fig:basic-sl}) laying on a Cartesian plane. In this section we will introduce the Cartesian transformation operator which returns the Cartesian coordinate of an opinion.

First of all, let us define the axes of the Cartesian system we will adopt. 

\begin{definition}
  Given the SL plane region \tri{BDU}, the \emph{associated Cartesian system} is composed by two axes, named respectively $x, y$, where the unit vector of the $x$ axis $\versor{x} = \frac{1}{|\vv{BD}|} \vv{BD}$, the unit vector of the $y$ axis $\versor{y} = \versor{\uncertainty{}}$, and $B$ is the origin.
\end{definition}

\begin{figure}[h]
  \centering
  \includegraphics[scale=0.7]{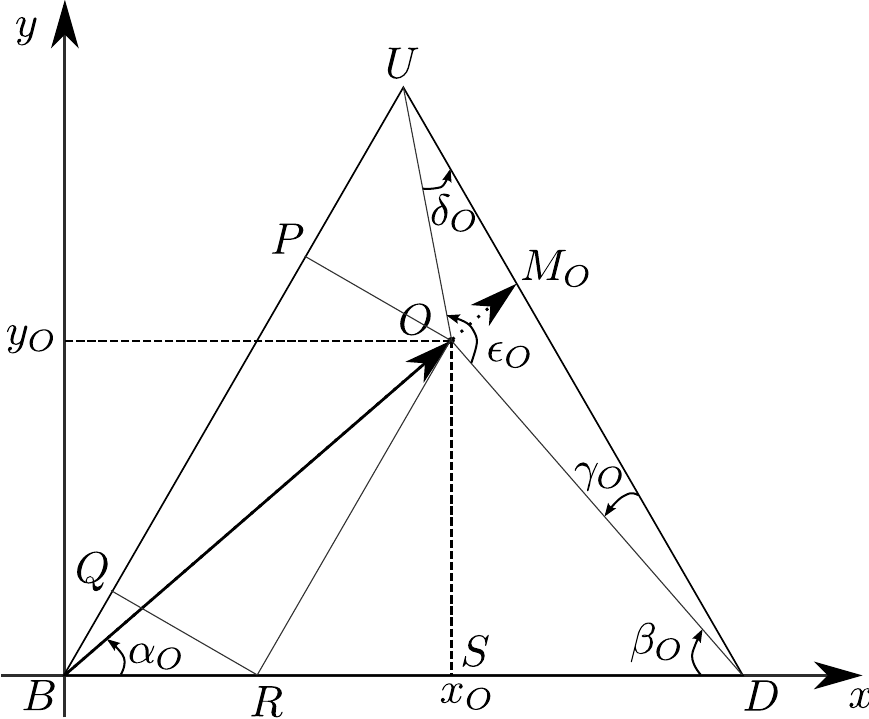}
  \caption{An opinion and its representation in the Cartesian system}
  \label{fig:cartesian-plane}
\end{figure}

Figure \ref{fig:cartesian-plane} depicts this Cartesian system.

The correspondence between the three values of an opinion and the corresponding coordinate in the Cartesian system we defined is shown in the following proposition.

\begin{proposition}
\label{propn:cartesian-transformation}
  Given a SL plane region \tri{BDU} and its associated Cartesian system $\tuple{x, y}$, an opinion $O \triangleq \opinion{O}$ is identified by the coordinate $\tuple{x_O, y_O}$ \st:
  \begin{itemize}
  \item $\displaystyle{x_O \triangleq \frac{\disbelief{O} + \uncertainty{O}~\cos(\frac{\pi}{3})}{\sin(\frac{\pi}{3})}}$;
  \item $y_O \triangleq \uncertainty{O}$.
  \end{itemize}

  \begin{proof}
    Proving that $y_O \triangleq \uncertainty{O}$ is trivial. 

    Let us focus on the first part of the proposition. Consider Figure \ref{fig:cartesian-plane}. Given $O$, we note that the for the point $P$,  $\frac{1}{|\vv{P O}|} \vv{P O} = \versor{b}$ (\ie{} $\vv{PO}$ is parallel to the disbelief axis) and $\frac{1}{|\vv{BP}|} \vv{BP} = \frac{1}{|\vv{BU}|} \vv{BU}$ (\ie{} $P$ is on the line $\vv{BP}$), and therefore $\ang{BPO} = \frac{\pi}{2}$. Then we must determine $Q$ and $R$ \st{} $\vv{Q R} = \vv{P O}$ and $y_{R} = 0$. By construction $|\vv{P O}| = |\vv{Q R}| = \disbelief{O}$, $\ang{Q R B} = \frac{\pi}{6}$, $\ang{ORD} = \frac{\pi}{3}$, and $x_O \triangleq |\vv{BS}| =  |\vv{BR}| + |\vv{RS}|$, where $|\vv{BR}| = \frac{\disbelief{O}}{\sin(\frac{\pi}{3})}$, and $|\vv{RS}| = \frac{\uncertainty{O}}{\sin(\frac{\pi}{3})} \cos(\frac{\pi}{3})$. 
  \end{proof}
\end{proposition}

There are some notable elements of Fig. \ref{fig:cartesian-plane} that we will repeatedly use below, and we therefore define them as follows:
\begin{itemize}
\item the angle $\alpha_O$ determined by the $x$ axis and the vector $\vv{BO}$;
\item the three angles ($\gamma_O, \delta_O,$ and $\epsilon_O$) of the triangle $\tri{ODU}$, namely the triangle determined by linking the point $O$ with the vertex $D$ and $U$ through straight lines.
\end{itemize}

\begin{definition}
  \label{def:properties}
  Given the SL plane region \tri{BDU}, given $O=\opinion{O}$ whose coordinates are $\tuple{x_O, y_O}$ where $\displaystyle{x_O \triangleq \frac{\disbelief{O} + \uncertainty{O}~\cos(\frac{\pi}{3})}{\sin(\frac{\pi}{3})}}$ and $y_O \triangleq \uncertainty{O}$, let us define and (via trivial trigonometric relations) compute the following.
  \begin{itemize}
  \item $\displaystyle{\alpha_O \triangleq \ang{OBD}} = $

    $\left\{
        \begin{array}{l l}
          0 & \mbox{if } \belief{O} = 1\\
          \displaystyle{\arctan\left(\frac{\uncertainty{O}~\sin(\frac{\pi}{3})}{\disbelief{O} + \uncertainty{O}~\cos(\frac{\pi}{3})}\right)} & \mbox{otherwise}\\
        \end{array} \right.$;
  \item $\displaystyle{\beta_O \triangleq \ang{ODB} } = $

    $\left\{
      \begin{array}{l l}
        \displaystyle{\frac{\pi}{3}} & \mbox{if } \disbelief{O} = 1\\
        \displaystyle{\arctan\left(\frac{\uncertainty{O}~\sin(\frac{\pi}{3})}{1-(\disbelief{O}+\uncertainty{O}~\cos(\frac{\pi}{3}))}\right)} & \mbox{otherwise}\\
      \end{array}\right.$;
  \item $\displaystyle{\gamma_O \triangleq \ang{ODU} = \frac{\pi}{3}} - \beta_O$;
  \item $\displaystyle{\delta_O \triangleq \ang{OUD} = }$

      $\left\{
        \begin{array}{l l}
          0 & \mbox{if } \uncertainty{O} = 1\\
          \displaystyle{\arcsin\left(\frac{\belief{O}}{|\vv{OU}|}\right)} & \mbox{otherwise}\\
        \end{array}\right.$;
  \item $\epsilon_O \triangleq \ang{DOU} = \pi - \gamma_O - \delta_O$;
  \end{itemize}

\noindent
where $\displaystyle{|\vv{OU}| = \sqrt{\frac{1}{3} (1 + \disbelief{O} - \uncertainty{O})^2 + \belief{O}^2}}$.

The angle $\alpha_O$ is called the  \emph{direction of $O$}.

Equivalently, we can  write $\vv{BO}$ or $\tuple{B, \alpha_O, |\vv{BO}|}$.

\end{definition}

Finally, as an element of SL is bounded to have its three components between $0$ and $1$, we are also interested in determining the point $M_O$ such that the vector $\vv{BM_O}$ has the maximum magnitude given (a) the direction $\alpha_O$ of an opinion $O$, and (b) $M_O$ is a SL opinion. In other words, determining the magnitude of $\vv{BM_O}$ will allow us to re-define the vector $\vv{BO}$ as a fraction  of $\vv{BM_O}$. 

\begin{definition}
  \label{defn:max-vector}
  Given the SL plane region \tri{BDU}, and $O \triangleq \opinion{O}$ whose coordinates are $\tuple{x_O, y_O}$ where $\displaystyle{x_O \triangleq \frac{\disbelief{O} + \uncertainty{O}~\cos(\frac{\pi}{3})}{\sin(\frac{\pi}{3})}}$ and $y_O \triangleq \uncertainty{O}$, and $\displaystyle{\alpha_O \triangleq \ang{OBD} =}$ $ \displaystyle{\arctan\left(\frac{\uncertainty{O}~\sin(\frac{\pi}{3})}{\disbelief{O} + \uncertainty{O}~\cos(\frac{\pi}{3})}\right)}$, let us define $M_O \triangleq \tuple{x_{M_O}, y_{M_O}}$ as the intersection of the straight line passing for $O$ and $B$, and the straight line passing for $U$ and $D$, and thus define the following.
  \begin{itemize}
  \item $\displaystyle{x_{M_O} \triangleq \frac{2 - y_O + \tan(\alpha_O)~ x_O}{\tan(\alpha_O) + \sqrt{3}}}$;

  \item $\displaystyle{y_{M_O} \triangleq -\sqrt{3}~ x_{M_O} + 2}$.
  \end{itemize}
\end{definition}

\subsection{The Trustworthiness-Confidence Combination Operator}

\begin{figure}[h]
  \centering
  \includegraphics[scale=0.8]{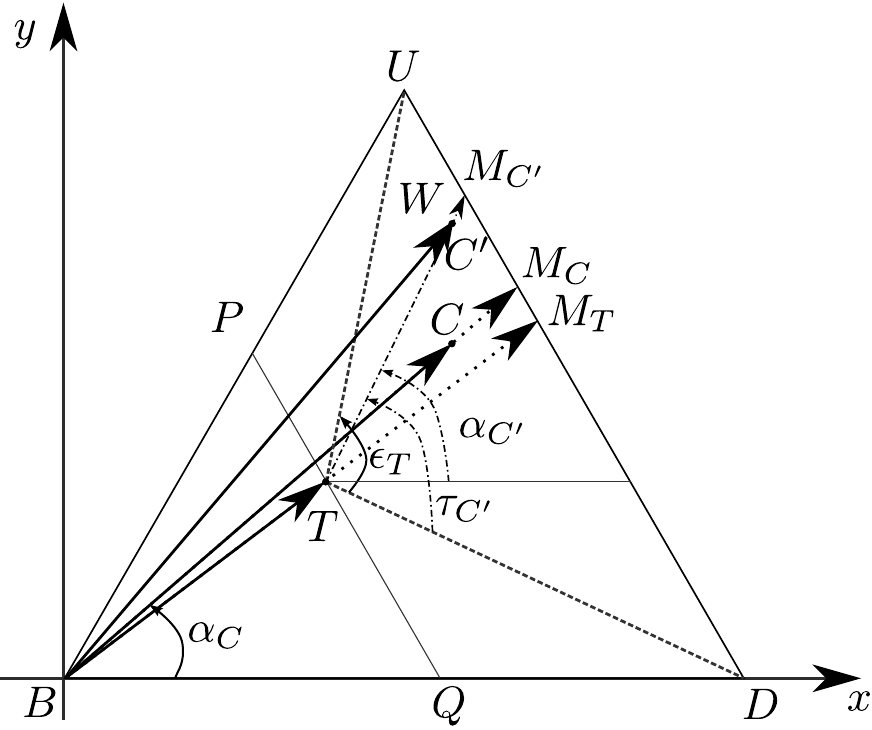}
  \caption{Projection of the confidence opinion and combination with the trustworthiness opinion}
  \label{fig:vector-confidence-combination}
\end{figure}

In this  paper, we  analyse a single method for projecting the confidence opinion $C$ into the space of SL opinions in order to satisfy the requirement that $\belief{W} \leq \belief{T}$. In particular, we  rely only on straight lines, leading to the definition of a new vector $\vv{TC'} = \tuple{T, \alpha_{C'}, |\vv{TC'}|}$ given the points \T{} and \C. Figure \ref{fig:vector-confidence-combination} depicts this situation. In particular, $\alpha_{C'}$ is such that the following proportion holds: $\alpha_C : \frac{\pi}{3} = \tau_{C'} : \epsilon_T$, where $\tau_{C'}$ is the angle $\ang{DTC'}$, and  $|\vv{TC'}|$ is such that the following proportion holds: $|\vv{BC}| : |\vv{BM_C}| = |\vv{TC'}| : |\vv{TM_{C'}}|$. $\alpha_{C'}$ represents the angle that the vector $\vv{TC'}$ makes to the vector parallel to the $x$ axis, and thus, from Definition \ref{def:properties}, we can write $\vv{TC'} = \tuple{T, \alpha_{C'}, |\vv{TC'}|}$. 

In other words, when $\alpha_C = 0$, then $\tau_{C'} = 0$ and thus we are projecting $C$ in the direction $\frac{1}{|\vv{TD}|} \vv{TD}$, while if $\alpha_C = \frac{\pi}{3}$ then we project $C$ in the direction $\frac{1}{|\vv{TU}} \vv{TU}$. It is important to note that due to our choice  of using straight lines to project $C$, the available space for projecting \C{} is reduced from the four-sided figure $UDQP$ (in Fig. \ref{fig:vector-confidence-combination}) to the triangle $\tri{TDU}$.

Given the new vector $\vv{TC'} = \tuple{T, \alpha_{C'}, |\vv{TC'}|}$, the result of the combination of the trustworthiness opinion $T$ and the confidence opinion $O$ is the opinion $W$ determined as the vector sum $\vv{BW} = \vv{BT} + \vv{TC'}$ in the Cartesian system $\tuple{x, y}$. This leads us to our core result.

\begin{definition}
\label{defn:combination}
  Given a trustworthiness opinion $T = \opinion{T}$ and a confidence opinion $C = \opinion{C}$, the combination of $T$ and $C$ is $W = \optrustconf{T}{C}$, where:
  \begin{itemize}
  \item $\uncertainty{W} = \uncertainty{T} + \sin(\alpha_{C'}) |\vv{TC'}|$;
  \item $\disbelief{W} = \disbelief{T} + (\uncertainty{T} - \uncertainty{W}) \cos(\frac{\pi}{3}) + \cos(\alpha_{C'}) \sin(\frac{\pi}{3}) |\vv{TC'}|$.
  \end{itemize}

  In particular:
  \begin{itemize}
  \item $\displaystyle{\alpha_{C'} = \frac{\alpha_C~\epsilon_T}{\frac{\pi}{3}}} - \beta_T$;
  \item $\displaystyle{|\vv{TC'}| = \frac{|\vv{BC}|}{|\vv{BM_{C}}|} |\vv{TM_{C'}}| =}$ \\ 
$=\displaystyle{r_C~|\vv{TM_{C'}}|}$
  \end{itemize}

  \noindent
  with $r_C = \frac{|\vv{BC}|}{|\vv{BM_C}|}$, and 

  \noindent
  $|\vv{TM_{C'}}| = \left\{
    \begin{array}{l l}
      2~\belief{T} & \mbox{if } \displaystyle{\alpha_{C'} = \frac{\pi}{2}}\\
      \displaystyle{\frac{2}{\sqrt{3}}~\uncertainty{T}} & \mbox{if } \displaystyle{\alpha_{C'} = - \frac{\pi}{3}}\\
      \displaystyle{\frac{2}{\sqrt{3}}~(1 - \uncertainty{T})} & \mbox{if } \displaystyle{\alpha_{C'} = \frac{2}{3} \pi}\\
      \displaystyle{\frac{2 \sqrt{\tan^2(\alpha_{C'}) + 1}}{| \tan(\alpha_{C'}) + \sqrt{3} | } ~\belief{T}} & \mbox{otherwise}\\
    \end{array}\right.$
\end{definition}

We now show that the combination requirements specified earlier are satisfied by $\optrustconf{}{}$.

\begin{theorem}
  Given $T = \opinion{T}$, $C = \opinion{C}$, and $W = \optrustconf{T}{C}$, then:
  \begin{enumerate}
  \item $W = \opinion{W}$ is an opinion;
  \item if $C = \tuple{1, 0, 0}$, then $W = T$;
  \item if $C = \tuple{0, 1, 0}$, then $W = C$;
  \item if $C = \tuple{0, 0, 1}$, then $W = C$;
  \item $\belief{W} \leq \belief{T}$.
  \end{enumerate}

  \begin{proof}
    Proving the thesis in the limit case is trivial. In the following we will assume, without loss of generality, that $\alpha_{C'} \neq \frac{\pi}{2}$, $\alpha_{C'} \neq -\frac{\pi}{3}$, $\alpha_{C'} \neq \frac{2}{3} \pi$.

    \textbf{Point 1.} $W = \opinion{W}$ must respect 
    \begin{equation}
      \label{eq:1}
      \uncertainty{W} + \disbelief{W} \leq 1
    \end{equation}

    From Def. \ref{defn:combination} it is clear that Equation \ref{eq:1} can be rewritten as follows.

    \begin{equation}
      \label{eq:2}
      \begin{split}
        & \uncertainty{T} + \disbelief{T} + \\ & + \frac{r_C}{2} \frac{2 \sqrt{\tan^2(\alpha_{C'}) + 1}}{|\tan(\alpha_{C'}) + \sqrt{3}|} ~\belief{T} ~(\sin(\alpha_{C'}) + \sqrt{3}\cos(\alpha_{C'})) \leq 1
      \end{split}
    \end{equation}
    
    In turn, using the relation $\tan(\alpha_{C'}) = \frac{\sin(\alpha_{C'})}{\cos(\alpha_{C'})}$, this can be rewritten as 

    \[
    \uncertainty{T} + \disbelief{T} + r_C \belief{T} \leq 1
    \]

    \noindent
    which entails the requirement that $r_C \leq \frac{1 - \uncertainty{T} - \disbelief{T}}{\belief{T}} = \frac{\belief{T}}{\belief{T}} = 1$. However, from definition \ref{defn:combination}, we know that $r_{C} \leq 1$, fulfilling this requirement.

    \textbf{Point 2.} $C = \tuple{1, 0, 0}$ implies that $|\vv{BC}| = 0$ and thus $r_C = 0$. Therefore, from Def. \ref{defn:combination}, $\uncertainty{W} = \uncertainty{T} + \sin(\alpha_{C'}) r_C |\vv{TM_{C'}}| = \uncertainty{T}$ and this results also implies that $\disbelief{W} = \disbelief{T}$. Since Point 1 shows that $W$ is an opinion in SL, we conclude that $W = T$.

    \textbf{Point 3.} $C = \tuple{0, 1, 0}$ implies that $r_C = 1$, and $\alpha_C = 0$ which, in turn, implies, from Definition \ref{defn:combination}, that $\alpha_{C'} = - \beta_T$. Also from Definition \ref{defn:combination}, $\uncertainty{W} = \uncertainty{T} - \sin(\beta_T) \frac{2 \sqrt{\tan^2(\beta_C) + 1}}{|-\tan(\beta_C) + \sqrt{3}|}$. Then, recalling Definition \ref{def:properties} and the trigonometric property that $\sin(\arctan(v)) = \frac{v}{\sqrt{1+v^2}}$, we obtain

    \begin{equation}
      \label{eq:3}
      \uncertainty{W} = \uncertainty{T} - \frac{\sqrt{3} \uncertainty{T}}{1 - (\disbelief{T} + \frac{\uncertainty{T}}{2})} \frac{1 - (\disbelief{T} + \frac{\uncertainty{T}}{2})}{\sqrt{3} (1 - \disbelief{T} - \uncertainty{T})} \belief{T} = 0
    \end{equation}

    Similarly, 
    \begin{equation}
      \label{eq:4}
      \begin{split}
        \disbelief{W} & = \disbelief{T} + \frac{\uncertainty{T}}{2} + \frac{1 - (\disbelief{T} + \frac{\uncertainty{T}}{2})}{\frac{1}{2} (1 - \disbelief{T} - \uncertainty{T})} \belief{T} = \\
        & = \disbelief{T} + \frac{\uncertainty{T}}{2} + 1 - \disbelief{T} - \frac{\uncertainty{T}}{2} = 1
      \end{split}
    \end{equation}

    Therefore, from Eq. \ref{eq:3} and Eq. \ref{eq:4}, and from Point 1, $W = \tuple{0, 1, 0} = C$.

    \textbf{Point 4.} $C = \tuple{0, 0, 1}$ implies $r_C = 1$, $\alpha_{C'} = \frac{\frac{\pi}{3} \epsilon_T}{\frac{\pi}{3}} - \beta_T = \epsilon_T - \beta_T = \frac{2}{3} \pi - \delta_t$. Therefore, we obtain that $\uncertainty{W} = \uncertainty{T} + \frac{\belief{T}}{2} (1 + \frac{\sqrt{3}}{\tan(\delta_T)})$. From Definition \ref{def:properties} and the trigonometric property that $\tan(\arcsin(v)) = \frac{v}{\sqrt{1 - v^2}}$ we obtain that $\uncertainty{W} = \uncertainty{T} + \frac{\belief{T}}{2} + \frac{\sqrt{3}}{2} \sqrt{|\vv{TU}|^2 - \belief{T}^2}$. From Definition \ref{def:properties} we can write:

    \begin{equation}
      \label{eq:5}
      \begin{split}
        \uncertainty{W} & = \uncertainty{T} + \frac{\belief{T}}{2} + \frac{\sqrt{3}}{2} \frac{1 + \disbelief{T} - \uncertainty{T}}{\sqrt{3}}\\
        & = \frac{1}{2} (1 + \belief{T} + \disbelief{T} + \uncertainty{T} ) = 1
      \end{split}
    \end{equation}

    Similarly, $\disbelief{W} = \disbelief{T} + \frac{\uncertainty{T}}{2} - \frac{1}{2} + \frac{\sqrt{3}}{2} \frac{1}{\sin(\delta_T)} \belief{T} = \disbelief{T} + \frac{\uncertainty{T}}{2} - \frac{1}{2} + \frac{\sqrt{3}}{2} |\vv{TU}|$. From Def. \ref{def:properties} we have

    \begin{equation}
      \label{eq:6}
      \begin{split}
        \disbelief{W} & = \disbelief{T} + \frac{\uncertainty{T}-1}{2} + \frac{3}{4}\belief{T} - \frac{\sqrt{3}}{4} \belief{T} \frac{1 + \disbelief{T} - \uncertainty{T}}{\sqrt{3} \belief{T}} \\ 
        & = \frac{1}{4} (4\disbelief{T} + 2 \uncertainty{T} - 2 + 3 \belief{T} - 1 + \uncertainty{T} - \disbelief{T}) = 0
      \end{split}
    \end{equation}

    From Equations \ref{eq:5} and \ref{eq:6}, together with Point 1, it follows that $W = \tuple{0,0,1}=C$.

    \textbf{Point 5.} Suppose instead $\belief{W} > \belief{T}$.

    \begin{equation*}
      \begin{split}
        & 1 - \disbelief{W} - \uncertainty{W} > 1 - \disbelief{T} - \uncertainty{T}\\
        & \disbelief{W} + \uncertainty{W} < \disbelief{T} + \uncertainty{T} \\
        & \disbelief{T} + \sin(\alpha_{C'} + \frac{\pi}{3}) \frac{r_C}{\sin(\alpha_{C'} + \frac{\pi}{3})} + \uncertainty{T} < \disbelief{T} + \uncertainty{T}\\
        & r_C < 0
      \end{split}
    \end{equation*}
    
    \noindent
    but $0 \leq r_C \leq 1$. \emph{Quod est absurdum.}
  \end{proof}
\end{theorem}

\section{Related and Future Works}
\label{sec:disc-future-works}

In this paper we describe an approach to computing a final level of trust given degrees of trustworthiness and confidence expressed as SL opinions. The main body of related work therefore relates to the calculation of degrees of confidence, as well as works detailing SL operators for opinion combination.

The first topic has received only limited attention in the literature.  In particular, \cite{Bhattacharya1998} describe --- in very general terms --- the importance of confidence in the process that leads to the definition of the degree of trust, but  do not provide any quantitative approach for combining the different components of trust.
More recently, \cite{Teacy2005,Huynh2006,Patel2007} discussed TRAVOS, a trust model that is built upon probability theory and based on observations of past interaction between agents. Moreover TRAVOS  calculates the confidence of its trust values given an acceptable level of error: if the confidence level of a trust value is below a predetermined minimum level, TRAVOS will
seek witness information about the target agent's past performance. TRAVOS in turn was extended in \cite{teacy12efficient}, but this fundamental mechanism remains unchanged. Critically, this notion of confidence is based on the number of previous interactions and level of reputational information obtained, rather than context as in the current work.

Interestingly, there are correlations between the level of confidence and the presence of obfuscated information. Investigations into the links between trust and such obfuscated data have only recently begun \cite{Bisdikian2012,Sensoy2012}, and to our knowledge, no approaches have been described for deriving a trust degree for a source of information which is obfuscating data.
 
Our work suggests several avenues of future work. First, we intend to investigate the overlaps between confidence obtained from systems such as TRAVOS and contextual confidence as described here. We will study how a degree of confidence can be derived from the context, and the differences with TRAVOS' notion of confidence. We also seek to determine whether our newly defined operator is beneficial in the context of reputational information and transitive trust. In order to do so, we intend to perform both an empirical and theoretical comparison between our system and existing ones. It is important to stress once more however that confidence, as expressed in existing work, does not deal with different contexts as we do in the current work.

Another avenue of future work will investigate how different levels of confidence can be used to reveal the presence of obfuscated information, improving the decision making process in such situations.

With regards to other SL operators, both the deduction and the discounting operator proposed in the literature \cite{Josang2006} bear some similarities to the operator proposed in this paper. The SL deduction operator considers three opinions, namely an opinion for an event $p$, an opinion for the derivation rule $p \rightarrow c$, and an opinion for the derivation rule $\lnot p \rightarrow c$, and deduces an opinion on the event $c$ by ``projecting''  the precedent opinion triangle in the subsequent sub-triangle (\cf{} \cite{Josang2005}). However, the deduction operator and the trustworthiness-confidence-combination operator we provide in this paper are not directly comparable, as J{\o}sang's deduction operates on three elements, while our operator is binary (this is also the reason why we did not consider deduction or abduction in Table \ref{tab:comparison-josang-operators}).

The discount operator was designed  to propagate trust degrees across a trust network. In particular, the situation in Example \ref{ex:scenario} where an agent $X$ is deciding whether or not to trust a source of information $Y$ can be rephrased as follows. An agent $X$ should decide whether or not to trust another agent $K$ which is a mediator between $X$ and the source of information $Y$ such that $K$ has his own degree of trust of $Y$. However, this situation is  different to  the one we examine --- a mediator acts either as a repeater, and in this case $X$ will always trust $K$, or is a fully-fledged agent, in which case $X$'s trust on $K$ about messages derived from $Y$ must consider the beliefs, intentions, and desires of $K$.

A deeper (and eventually empirical too) evaluation of the differences between J{\o}sang discount and deduction operators therefore forms another important strand of envisioned future work. In addition, we intend to analyse these and other SL operators and properties from a geometrical perspective.

Finally, we are currently working on providing different ways for projecting the confidence opinion into the set of SL opinions satisfying the fourth combination requirement. As Figure \ref{fig:vector-confidence-combination} depicts, this will lead (in general) to a four-sided figure, unlike the current work, where  the projected opinion is always inside a triangle which is strictly contained within that four-side figure. To do so, we will consider non-linear projection methods in the future, as opposed to the simple linear projection method applied in the current work.

\section{Conclusions}
\label{sec:conclusions}

In  this paper we introduced an approach for deriving a trust degree on a source of information given a degree of trustworthiness and a confidence degree, with the latter potentially being  determined by the current interaction context. In particular, the trustworthiness degree can be derived from past interactions with the source of information, which leads to a SL opinion. By expressing confidence as a SL opinion, we are able to describe an approach for merging the two degrees, this being one of our  main contributions. Critically, we have shown that none of the existing SL operators satisfy the requirements necessary to perform this merging. Since our proposed operator is based on geometrical properties, we also  introduce a formal geometric method for analysing SL opinions. While the contributions of this paper are clearly important, this work opens up a rich direction for future work, which we have already begun actively pursuing.

\section*{Acknowledgements}
Research was sponsored by US Army Research laboratory and the UK Ministry of Defence and was accomplished under Agreement Number W911NF-06-3-0001. The views and conclusions contained in this document are those of the authors and should not be interpreted as representing the official policies, either expressed or implied, of the US Army Research Laboratory, the U.S. Government, the UK Ministry of Defense, or the UK Government. The US and UK Governments are authorized to reproduce and distribute reprints for Government purposes notwithstanding any copyright notation hereon.
 
 
 
 


\bibliographystyle{abbrv}
\bibliography{biblio} 
  
\end{document}